\documentclass{jfm}
\usepackage[OT1]{fontenc} 
\usepackage{graphicx}
\usepackage{float}
\usepackage{epstopdf, epsfig}
\usepackage{graphicx}
\usepackage{subfigure}
\usepackage{color}
\usepackage{psfrag}
\usepackage{natbib}
\usepackage{multirow}
\usepackage{psfrag}
\usepackage{txfonts}
\usepackage{xfrac}
\usepackage{sidecap}
\usepackage{booktabs}

\usepackage{enumitem}
\usepackage{pstricks}
\usepackage{psfrag}
\usepackage{tikz}
\usepackage{txfonts}

\usepackage{calc}

\usepackage{caption}
\usepackage{rotating}
\usepackage{booktabs}
\usepackage{multirow}
\usepackage{tabularx}
\usepackage{upgreek}
\usepackage[normalem]{ulem}

\usepackage{bm}
\usepackage{mathrsfs}

\definecolor{green_m}{rgb}{0,0.5,0}
\definecolor{magenta_m}{rgb}{1,0,1}
\definecolor{magenta1}{rgb}{.75,0,0.75}

\begin{document}

\author{Mogeng Li\aff{1}, Charitha M. de Silva\aff{1},  Amirreza Rouhi\aff{1}, Rio Baidya\aff{1}, Daniel Chung\aff{1}, Ivan Marusic\aff{1} and Nicholas Hutchins\aff{1}}

\title{\color{black} Recovery of the wall-shear stress to equilibrium flow conditions after a rough-to-smooth step-change in turbulent boundary layers}

\shorttitle{Recovery of the wall-shear stress after a R$\rightarrow$S change in TBL}
\shortauthor{M. Li and others}

\affiliation{\aff{1}Department of Mechanical Engineering, University of Melbourne, Victoria 3010, Australia}

\date{Received: date / Accepted: date}

\maketitle

\begin{abstract}

This paper examines recovery of the wall-shear stress of a turbulent boundary layer that has undergone a sudden transition from a rough to a smooth surface. Early works of \cite{antonia1972response} questioned the reliability of standard smooth-wall methods to measure wall-shear stress in such conditions, and subsequent studies show significant disagreement depending on the approach used to determine the wall-shear stress downstream. Here we address this by utilising a collection of experimental databases at $Re_\tau \approx 4100$ that have access to both `direct' and `indirect' measures of the wall-shear stress to understand the recovery to equilibrium conditions to the new surface.  Our results reveal that the viscous region ($z^+\lesssim 4$) recovers almost immediately to an equilibrium state with the new wall conditions, however, the buffer region and beyond takes several boundary layer thicknesses before recovering to equilibrium conditions, which is longer than previously thought. A unique direct numerical simulation database of a wall-bounded flow with a rough-to-smooth wall transition is employed to confirm these findings. In doing so, we present evidence that any estimate of the wall-shear stress from the mean velocity profile in the buffer region or further away from the wall tends to underestimate its magnitude in the near vicinity of the rough-to-smooth transition, and this is likely to be partly responsible for the large scatter of recovery lengths to equilibrium conditions reported in the literature. Our results also reveal that the smaller energetic scales in the near-wall region recover to an equilibrium state associated with the new wall conditions within one boundary layer thickness downstream of the transition, while the larger energetic scales exhibit an over-energised state for several boundary layer thicknesses downstream of the transition. Based on these observations, an alternative approach to estimating the wall-shear stress from the premultiplied energy spectrum is proposed.

\end{abstract}


\section{Introduction}

\label{sec:intro}

\begin{figure}
\vspace{3mm}
\centering
\raisebox{9mm}[0pt][0pt]{%
\makebox[0.5\textwidth][c]{\includegraphics[trim = 0 0mm 0 0,scale = 0.45,clip]{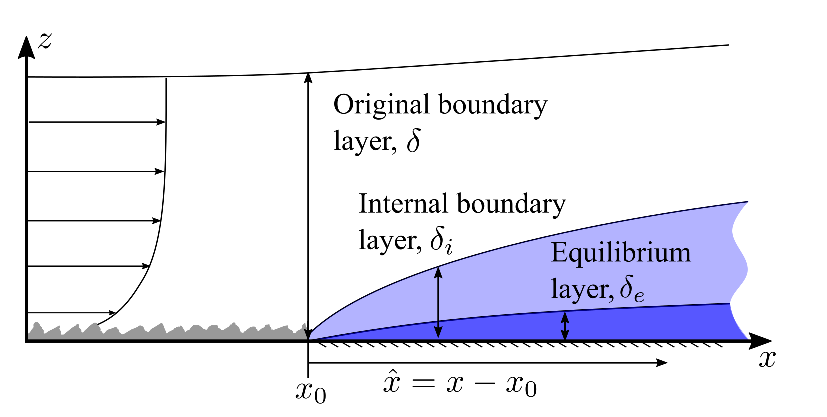}}}
\includegraphics[trim = 0 0mm 0 0,scale = 0.8,clip]{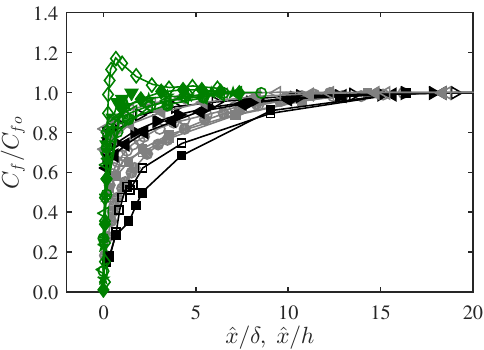}  
 \put(-390,135){(\textit{a})}  \put(-190,135){(\textit{b})} 
\caption{(\textit{a}) Schematic of a turbulent boundary layer flow over a rough-to-smooth change in surface condition. Flow is from left to right and $\hat{x} = x-x_0$ represents the fetch measured from the rough-to-smooth transition which occurs at $x=x_0$. (\textit{b}) Recovery of skin friction coefficient $C_f$ recovery downstream of a rough-to-smooth transition from a range of numerical and experimental databases. Details of each dataset are summarised in table \ref{tab:DataSumm}. The colours of the symbols indicate the region where $C_f$ is determined: green represents $C_f$ measured within the viscous sublayer or directly at the wall, grey represents buffer layer and black the logarithmic layer.  Results are normalised  by $C_{fo}$, which corresponds to the most downstream reported $C_f$ measurement from each dataset. }
\label{fig:IBL_sketch}
\end{figure}

Surface roughness with heterogeneity is present in wall-bounded turbulent flows in a variety of conditions. For example,  the patchiness of biofouling on the hull of a ship or the changes in the surface roughness conditions that occur at the interface between forest and grasslands. Though such heterogeneity can occur in a wide range of configurations, one simple distillation of this problem is to consider a sudden transition from a rough-to-smooth surface occurring in the streamwise direction, as examined in the seminal work of \citet{antonia1972response}. This configuration is best described with reference to figure \ref{fig:IBL_sketch}(\textit{a}), where upstream of the transition, an equilibrium rough wall boundary layer has developed over the rough fetch. Following the transition, the new smooth wall condition initially modifies the near-wall region. The effect of the new wall condition then gradually propagates towards the interior of the flow with increasing distance downstream of the transition. The layer that separates the modified near-wall region (which `sees' the new smooth wall condition) from the unaffected oncoming flow, further away from the wall (which `remembers' the rough-wall condition) is generally referred to as the internal boundary layer (IBL) with a thickness denoted by $\delta_i$. The layer where the flow is in equilibrium with the new wall condition is referred to as the equilibrium layer \citep{Garratt1990, savelyev2005internal} with thickness $\delta_e$. In most cases, the majority of the flow within the IBL is still in non-equilibrium with the local wall condition, and a general consensus is $\delta_e$ takes up about $5\%$ of $\delta_i$ defined based on the shear stress profile adjustment for a flow over rough-to-smooth change \citep[see][]{Rao1974,Shir1972}. 

Although the streamwise rough-to-smooth heterogeneity has been studied extensively over the past few decades  \citep{Bradley1968,antonia1972response,Shir1972,Rao1974,Chamorro2009,Hanson2016}, to date, the recovery to equilibrium conditions of the new surface following such a transition is far from understood. For example, determining the local wall-shear stress  $\tau_w$ after the transition (and subsequently the friction velocity $U_{\tau}$) have been hampered by reliability issues.  \citet{antonia1972response} used three different techniques to determine $\tau_w$ following a rough-to-smooth transition (Clauser chart, Preston tube, and the momentum integral equation) noting that, \textit{`... none of the standard smooth-wall methods of obtaining skin friction from the mean profile is reliable for some distance downstream from the roughness change'.}

\begin{table}
\centering
\begin{tabular}{p{.5in}cp{0.5in}p{0.4in}p{0.6in}p{0.5in}cccp{0.3in}}                             
      Reference & Symbol & Technique & Region & Geometry& Roughness& $Re_{\tau}$ & $k_s^+$ & $\delta/k_p, \;h/k_p$ & $\Updelta H/k_p$ \\
      \hline
    \multirow{2}{.5in}{\cite{Hanson2016}} & {\color{gray}$\medcirc$} & \multirow{2}{0.5in}{Preston tube} & \multirow{2}{0.4in}{buffer} & \multirow{2}{.7in}{boundary layer} & P16 grit & $2.3\times10^3$ & $209$ & 29 & \multirow{2}{0.3in}{N/A}\\
     & {\color{gray}$\medbullet$} & &   &   & mesh & $3.5\times10^3$ & $1300$ & 19  \\
\\
\\
\\
\hline
    \multirow{4}{.5in}{\cite{antonia1972response}} & {\color{black}$\square$} & \multirow{2}{0.5in}{Clauser chart} & log & \multirow{4}{.7in}{boundary layer} & \multirow{4}{0.5in}{square ribs}& $3.4\times10^3$ & - & 22& 0\\
    & {\color{black}$\blacksquare$} &  & log &  & & $6.1\times10^3$ & -& 22& 0\\
    & {\color{gray}$\square$} & \multirow{2}{0.5in}{Preston tube} & buffer &  &  & $3.4\times10^3$ & - & 22& 0\\
    & {\color{gray}$\blacksquare$} &  & buffer &  &  & $6.1\times10^3$ & - & 22& 0\\
\hline
    \multirow{5}{.5in}{\cite{saito2014}} & {\color{gray}$\lhd$} & \multirow{5}{0.7in}{wall-modelled LES} & buffer & \multirow{5}{.7in}{full channel (both sides roughened)}  & \multirow{5}{0.5in}{modelled} & $2.1\times10^4$ & $21$ & \multirow{5}*{N/A}& 0\\
    & {\color{gray}$\blacktriangleleft$} &   & buffer &   &  & $2.2\times10^5$ & $219$ &  & 0\\
    & {\color{black}$\blacktriangleleft$} &   & log &   &  & $2.5\times10^6$ & $2458$ &  & 0\\
    & {\color{black}$\rhd$} &   & log &   &  & $2.3\times10^6$ & $225$ &  & 0\\
    & {\color{black}$\blacktriangleright$} &   & log &   &  & $2.4\times10^6$ & $1193$ &  & 0\\
\hline
    \multirow{3}{.5in}{\cite{ismail2018simulations}} & {\color{green_m} $\lozenge$} & \multirow{3}{0.7in}{DNS} & \multirow{3}{0.4in}{wall} & \multirow{3}{.7in}{full channel (one side roughened)} & \multirow{3}{0.5in}{square ribs} & $5.0\times10^2$ & 343 &12 & $-1$\\
    & {\color{green_m} $\lhd$} &   &   &  &   & $2.2\times10^3$ & 1540 &12& $-1$\\
    & {\color{green_m} $\star$} &   &   &  &   & $2.1\times10^3$ & 1105 &16& $-1$\\
    & {\color{green_m} $\blacklozenge$} &   &   &  &   & $2.4\times10^3$ & 2105&9.6& $-1$\\
\hline

\multirow{2}{.5in}{\cite{ismail2018effect}} & {\color{green_m} $\medcirc$} & \multirow{2}{0.7in}{DNS} & \multirow{2}{0.4in}{wall} & \multirow{2}{.7in}{full channel (one side roughened)} & \multirow{2}{0.5in}{3D cubes} & $1.6\times10^3$ & 338 &12 & $-1$\\
    & {\color{green_m} $\medbullet$} &   &   &  &   & $1.5\times10^3$ & 233 &12& $-1$\\
\\
\hline

    \cite{Chamorro2009} & {\color{green_m}$\blacktriangledown$}&near-wall hotwire& viscous & boundary layer & mesh & $1.5\times10^4$& $479$& $133$ & $-0.5$\\
\hline
 \end{tabular}
 \captionof{table}{Summary of published data on wall-shear stress recovery in wall-bounded flows downstream of a rough-to-smooth transition. $Re_{\tau}$ and $k_s^+$ correspond to the friction Reynolds number and the equivalent sand grain roughness Reynolds number at the rough-to-smooth transition, and $k_p$ is the maximum roughness height between the crest and trough. $\Updelta H$ is the height increment from the roughness crest to the smooth surface downstream. $\Updelta H = 0$ implies that the smooth wall is aligned with the roughness crest, whereas $\Updelta H<0$ indicates that the smooth wall is below the roughness crest.}
 \label{tab:DataSumm}
\end{table}

The scatter in the recovery of the wall-shear stress is highlighted in figure \ref{fig:IBL_sketch}(\textit{b}), which collates the skin friction coefficient, $C_f=\tau_w/(\frac{1}{2}\rho U^2_{\infty})$ (where $U_\infty$ is the freestream velocity and $\rho$ is the air density),  from a collection of experimental  and numerical databases (see table \ref{tab:DataSumm} for a summary of key parameters of each database). This leads to an uncertainty in drag over $0<\hat{x}/\delta<10$ ($\delta$ is defined as the $z$ location where $\overline{U} = 0.99 U_{\infty}$ at the roughness transition), defined as $C_{D10} \equiv \int_0^{10} C_f\,\mathrm{d}(\hat{x}/\delta)$, up to 40\%. In part, the disagreement between the databases is due to differences in Reynolds numbers, flow geometry (internal versus external wall-bounded flows) and surface conditions (the magnitude and type of the roughness change at $x = x_0$), however here we demonstrate that the method of $C_f$ measurement can introduce systematic variation in the reported $C_f$ recovery. The colours of the symbols in figure \ref{fig:IBL_sketch}(\textit{b}) broadly distinguish the data based on the $C_f$ measurement technique, with the black symbols showing methods operating in the log-region, the grey symbols indicating the buffer layer, and the green symbols showing measurements from the viscous sublayer (OFI, near-wall gradient, DNS etc). We conjecture that when $\tau_w$ is estimated directly at the wall or from the viscous region, a higher $C_f$, or a faster recovery is observed in general. Conversely, when inferred from velocity signals further away from the wall, lower $C_f$ and longer recovery lengths are observed. It is worth noting that similar challenges can arise when estimating $\tau_w$ in numerical studies. For example, many works bypass the expense of simulating changing surface conditions directly (using DNS) through the use of Reynolds-Averaged Navier Stokes~\citep{Rao1974} or Wall-Modelled Large-Eddy Simulations~\citep{bou2004large,saito2014,lopes2015determination}. In both approaches, the near-wall turbulence (below the logarithmic region) is inferred from modelling assumptions, which may not be applicable for flows in non-equilibrium conditions.  Here we use both experimental and numerical databases to provide evidence for the reasoning above and explain why different methods of measuring $C_f$ can explain some of the discrepancies in the literature.

 Throughout this paper, the coordinate system $x$, $y$ and $z$ refer to the streamwise, spanwise and wall-normal directions, respectively. The rough-to-smooth transition occurs at $x = x_0$, and we use the definition $\hat{x} = x-x_0$ for the fetch on the smooth wall downstream.  Corresponding instantaneous streamwise, spanwise and wall-normal velocities are represented by $\widetilde{U}$, $\widetilde{V}$ and $\widetilde{W}$, respectively, with velocity fluctuations given by lower case letters. Overbars indicate spanwise- and/or time-averaged quantities and the superscript $+$ refers to normalisation by local inner scales with $U_{\tau} = U_{\tau}(\hat{x})$.  For example, we use $l^{+} = lU_{\tau}/\nu$ for length and $\widetilde{U}^{+} = \widetilde{U}/U_{\tau}$ for velocity, where $U_{\tau}$ is the friction velocity and $\nu$ is the kinematic viscosity of the fluid.

\section{Experimental databases} \label{sec:exp_num_databases}

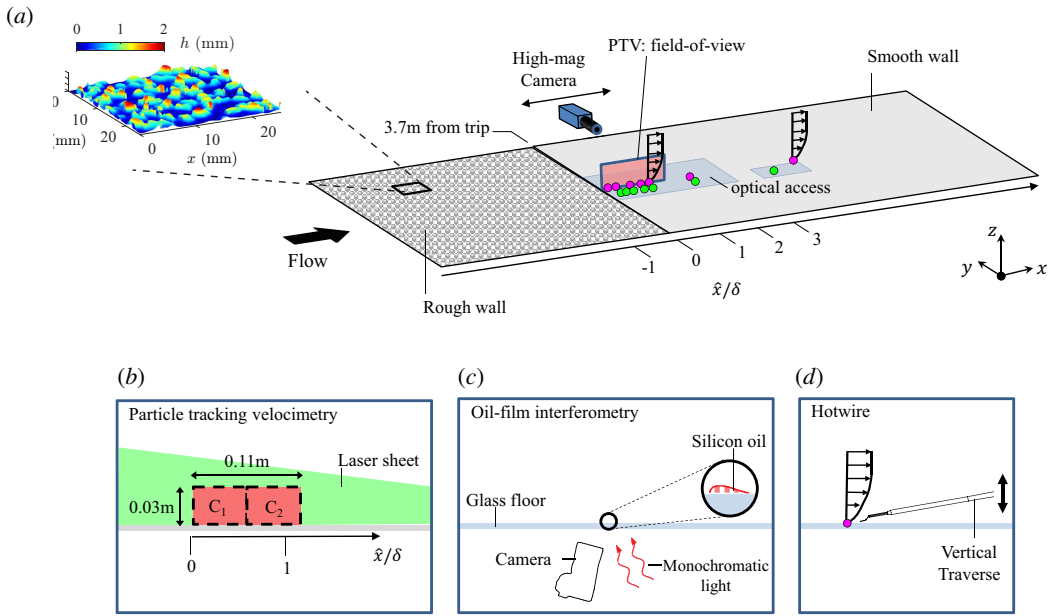
\begin{figure}
\centering
\includegraphics[trim = 5mm 15mm 0 15mm,scale = 0.52,clip]{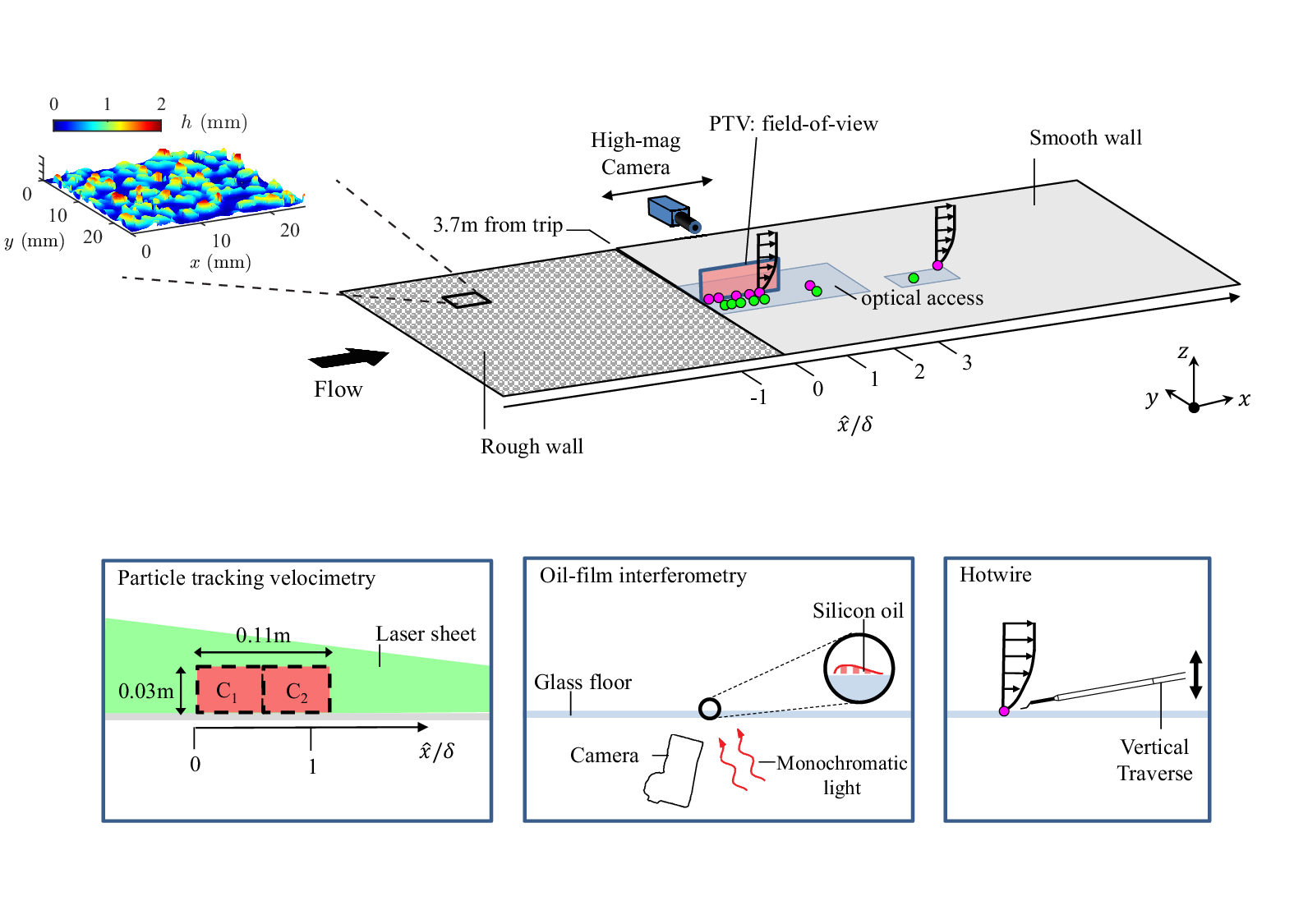}   \put(-405,230){(\textit{a})}   \put(-363,95){(\textit{b})}   \put(-235,95){(\textit{c})}   \put(-108,95){(\textit{d})} 
\caption{ (\textit{a}) Overview of  the experimental campaign in the open-return boundary layer wind tunnel facility at $Re_\tau \approx 4100$.  The {\color{magenta_m}$\medbullet$} symbols correspond to the locations of the hotwire wall-normal profiles, and the {\color{green}$\medbullet$} symbols correspond to the locations of OFI measurements.  The colour contours in (\textit{a}) illustrate the topography of the P16 grit sandpaper which is employed as the rough-walled surface and  (\textit{b-d}) illustrates the  particle tracking velocimetry,  oil-film interferometry and the hotwire traverse system, respectively. $C_1$ and $C_2$ in (\textit{b}) correspond to the field-of-views captured non-simultaneously using a scientific double-frame camera with a vertical laser sheet that is projected upstream through the working section. }
\label{fig:EXP_setup}
\end{figure}

The current experimental databases are acquired in an open-return boundary layer wind tunnel facility in the Walter Basset Aerodynamics Laboratory at the University of Melbourne.  The readers are referred to \citet{Marusic1995}, \citet{harun2013pressure}, \citet{nugroho2013large} and \citet{Kevin_2015} for further details of this facility. The turbulent boundary layer is tripped by a strip of P40 sandpaper at the inlet of the working section, and then develops on the tunnel floor. 
The arrangement of the experimental campaign consisting of hotwire boundary layer traverses, oil film interferometry measurements and PTV measurements are depicted in figure \ref{fig:EXP_setup}. For the present work,  the first 3.7 m of the tunnel surface is covered by P16 grit sandpaper, while the remaining streamwise length (1.9 m) is a smooth surface. Details of the roughness parameters are obtained by scanning a $60 \;\mathrm{mm} \times 60 \;\mathrm{mm}$ section of the sandpaper using an in-house built laser scanner. The maximum roughness height between the crest and trough is $k_p\approx2$ mm, which is equivalently 2\% of the boundary layer thickness $\delta$ at the surface transition. The roughness crest is approximately $3$ mm above the smooth surface, corresponding to $\Updelta H/k_p = -1.5$. The root-mean-squared roughness height ($k_{rms}$) of this surface is 0.387 mm, and the equivalent sandgrain roughness is $k_s \approx 2.8$ mm, yielding $k_s^+ \approx 130$ at $x_0$. The distribution of pressure coefficient $C_p\equiv(p-p_{ref})/\frac{1}{2}\rho U_{\infty}^2$ over the working section is obtained using static pressure taps mounted on the tunnel roof, and $C_p = 0\pm 0.01$ is achieved in most areas with no distinguishable localised pressure gradient observed in the vicinity of the rough-to-smooth change. The range of $C_p$ variation is comparable with other zero-pressure gradient studies conducted in the same facility \citep[see][]{harun2013pressure,Nugroho2015}, thus the pressure gradient effect over the current rough-to-smooth surface can be considered negligible. All measurements are acquired at a nominal freestream velocity of $U_\infty \approx 15$ m/s. The friction Reynolds number ($Re_\tau=\delta U_\tau /\nu$) immediately upstream of the transition location over the rough surface is $Re_\tau \approx 4100$.

\subsection{Hotwire anemometry} \label{section:HW}

A conventional single-wire hotwire probe of $2.5~\upmu \mathrm{m}$ diameter is operated by an in-house Melbourne University Constant Temperature Anemometer (MUCTA). Calibration is performed following an \textit{in-situ} procedure before and after each measurement. Thereafter, any drift is corrected by an intermediate single point re-calibration (ISPR) method discussed in \citet{talluru2014calibration}, where the hotwire voltage is periodically monitored in the freestream. The uncertainty in $U$ and $\overline{u^2}$ is usually within $1\%$ and $3\%$, respectively \citep{yavuzkurt1984guide}. The method of calibration drift correction proposed by \citet{talluru2014calibration} employed here offers further improvements. Boundary layer profiles  are taken at  $ \hat{x} =$ 10, 30, 60, 90, 180, 360 and 1190 mm, corresponding to $ \hat{x}/\delta = 0.11, 0.34, 0.68, 1.0, 2.0, 4.1$ and 13.4  ({\color{magenta_m}$\medbullet$} symbols in figure \ref{fig:EXP_setup}). Each profile consists of 50 logarithmically spaced measurement locations in the wall-normal direction for $0.4 \mathrm{mm} \lesssim z \lesssim 2\delta$, and the voltage signal is sampled at 30 kHz for 150 seconds at each wall-normal location, corresponding to a sample interval $\Updelta t^+ < 0.6$ and a total sampling duration $T_{\mathrm{samp}}$ of $2.25\times10^4$ boundary-layer turnovers ($T_{\mathrm{samp}} U_{\infty}/\delta$).

\subsection{Particle tracking velocimetry} \label{section:PTV}

To complement the hotwire databases with near-wall information, high magnification PTV measurements are performed immediately downstream of the rough-to-smooth transition. The magnified field of view targeted at the near wall region is ideally suited for this analysis, providing access to well-resolved velocity signals within the viscous sublayer region of the flow ($z^+\lesssim4$). A field of view of $1.2\delta \times 0.3 \delta$ is achieved by stitching two non-simultaneous measurements obtained at different streamwise locations C$_1$ and C$_2$, as shown in the inset of figure \ref{fig:EXP_setup}. A calibration target that spans the entire extent of the FOV, which has been proven to work well for multi-camera large-FOV experiments \citep[see][]{Silva2014a}, is employed to stitch the time-averaged statistics from the different camera positions together and also to account for image distortions. The uncertainty in the calibration of the pixel size in the current PIV/PTV measurement is approximately $0.6\%$, leading to a variation of $1.2\%$ in $\tau_w$.

The experimental image-pairs are processed using an in-house PIV/PTV package developed at the University of Melbourne \citep{Silva2014a}. To enhance the near-wall resolution, a hybrid PIV-PTV algorithm \citep{cowen1997hybrid} is used with $128 \times 8$ (75\% overlap) and $4 \times 4$ pixel integration window for the PIV and PTV pass, respectively. The wall-normal location of the PTV database is refined to subpixel accuracy by correlating the near wall particles and their reflections on a frame-by-frame basis.

\subsection{Oil film interferometry}

The wall-shear stress, $\tau_w$, is measured using oil film interferometry (OFI) \citep{zanoun2003evaluating, fernholz1996new}. The experimental configuration is illustrated in figure  \ref{fig:EXP_setup}(\textit{c}). A silicon oil droplet is placed on a clear glass surface and illuminated by a monochromatic light source from a sodium lamp. The resulting interference pattern is captured using a Nikon D800 DSLR camera. In a similar fashion to the PTV measurements, the FOV of the OFI measurements is calibrated with a  calibration grid featuring a 2.5 mm dot spacing, providing a conversion from image to physical space. 

For each OFI database, 100 images are captured with a time interval of five seconds between images. The image sequences are then processed using an FFT based algorithm \citep{ng2007oil}  to extract the fringe spacing of the interferograms. 
Thereafter, a linear trend is fitted to the extracted fringe spacing of the interferograms versus time to evaluate $\tau_w$. The main sources of uncertainty in the current OFI measurement lie in the calibration of oil viscosity and the camera calibration, and the relative error in the oil viscosity $\nu$ and the pixel size is estimated to be $0.5\%$ and $0.6\%$, respectively. Overall, considering other uncertainties associated with the fringe extraction and dust contamination of the oil film, the repeatability in $\tau_w$ obtained by OFI in the current study is estimated to be $\pm1.5\%$.

\section{Experimental results} \label{sec:exp_results}

\begin{figure}
\vspace{5mm}
\centering
\includegraphics[trim = 3 0mm 0 0,scale = 0.7,clip]{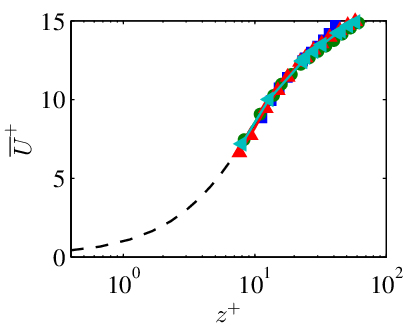}  \put(-130,112){(\textit{a})}  \put(-115,112){Hotwire - buffer fit} 
\includegraphics[trim = 7mm 0mm 0 0,scale = 0.7,clip]{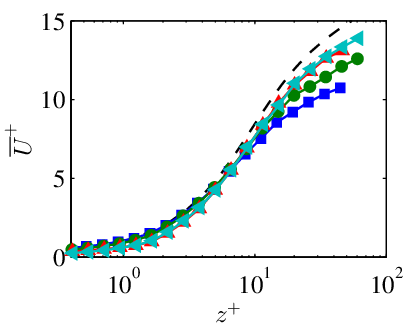}  \put(-130,112){(\textit{b})}  \put(-115,112){PTV - viscous sublayer fit} 
\includegraphics[trim = 7mm 0mm 0 0,scale = 0.7,clip]{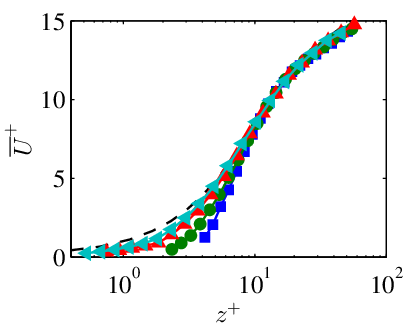}  \put(-130,112){(\textit{c})}  \put(-115,112){PTV - buffer fit}
\caption{ Mean streamwise velocity statistics from hotwire and PTV experimental data at $Re_\tau \approx 4100$. Normalisation is by friction velocity, $U_\tau$, estimated from a fit to the  (\textit{a,c}) buffer and  (\textit{b}) viscous sublayer regions, and $\hat{x}$ corresponds to the streamwise distance from the  rough-to-smooth transition. The black dashed line corresponds to a reference smooth-walled boundary layer DNS database at $Re_\tau  \approx 2500$ \citep{Sillero2013} and the {\color{blue}$-\blacksquare-$}, {\color{green_m}$-\!\!\medbullet\!\!-$}, {\color{red}$-\!\blacktriangle\!-$} and {\color{cyan}$-\!\!\blacktriangleleft\!\!-$} symbols correspond to results at $\hat{x}/\delta=$ 0.11, 0.34, 0.68 and 1.0, respectively.}
\label{fig:Umean_HW_PTV}
\end{figure}
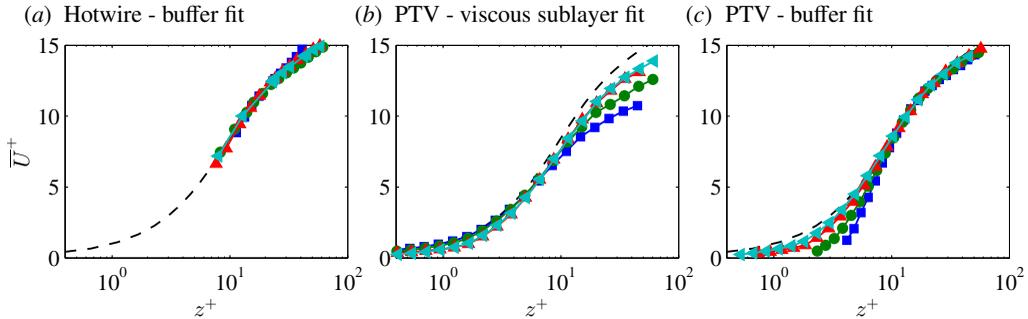

Figure \ref{fig:Umean_HW_PTV}(\textit{a}) shows the hotwire measured mean streamwise velocity profiles $\overline{U}$ at various locations downstream of the rough-to-smooth transition. Due to an inability to make hotwire measurements in the viscous sublayer, the friction velocity $U_\tau(\hat{x})$ for this figure has initially been estimated from a least squares fit in the buffer region ($10\lesssim z^+ \lesssim 30$) to a reference smooth-walled DNS profile from \citet{Sillero2013}. We note, due to uncertainty associated with the precise wall-normal location of the hotwire measurements, a wall-normal shift is included as a free parameter in the fit (the wall correction returned by the fit is typically within $0.35 \mathrm{mm}$). As dictated by the fit, the profiles in figure  \ref{fig:Umean_HW_PTV}(\textit{a}) exhibit an excellent collapse in the buffer region ($10\lesssim z^+ \lesssim 30$)  to the canonical case (dashed line). Critically, however, the quality of agreement in the near-wall region cannot be assessed due to the lack of near-wall data from the hotwire measurements.

To overcome this shortcoming, figure \ref{fig:Umean_HW_PTV}(\textit{b}) presents $\overline{U}$ from the PTV database, where a more direct estimate of  $\tau_w$  (hence $U_\tau$)  is accessible as we are able to compute  ${U}_{\tau}$   using a least squares fit in the viscous sublayer ($z^+ \lesssim 4$)  following
\begin{eqnarray} \label{eqn:utau_sublayer}
     {U}_{\tau}& = &\sqrt{\frac{\tau_{w}}{\rho}}  = \sqrt{\nu \frac{\partial \overline{U}}{\partial z}}.
\end{eqnarray}
Scaled in this way, the PTV data must exhibit collapse in the near wall region, and figure  \ref{fig:Umean_HW_PTV}(\textit{b})  shows a growing departure from the reference smooth wall profile with increasing $z$ in the buffer region ($10\lesssim z^+ \lesssim 30$), demonstrating that $U_\tau$  (and hence $\tau_{w}$) estimated from the buffer region and viscous sublayer region differ substantially. It should be noted that $U_\tau$ obtained from the near-wall gradient of the PTV data is believed to represent the correct estimate and matches very closely the value measured by oil-film interferometry (to be detailed further in figure \ref{fig:Cf_HW_PTV}). If we ascribe greater confidence to the PTV measured $U_\tau$, the most likely interpretation here is that the buffer region is yet to recover to an equilibrium state to the new surface conditions, and consequently underestimates the wall-shear stress immediately downstream of a rough-to-smooth transition. 

To further illustrate this behaviour, figure \ref{fig:Umean_HW_PTV}(\textit{c})  shows $\overline{U}$ from the PTV database normalised by a $U_\tau$ estimated from the buffer region (following the same procedure as applied to the hotwire data in figure \ref{fig:Umean_HW_PTV}(\textit{a}) ). Despite the subpixel accuracy in the wall location for the PTV data, a free parameter accounting for the wall-normal shift is also included in the fit to fully replicate the hotwire procedure. The results reveal a lack of agreement in the near wall region below $z^+ \lesssim 10$, which confirms that the collapse observed in figure \ref{fig:Umean_HW_PTV}(\textit{a,c})  in the buffer region ($10\lesssim z^+ \lesssim 30$) appears to be an artefact of an erroneous $U_\tau$.  It should also be noted that the estimated internal boundary layer thickness $\delta_i$ is $O(100)$ viscous units above the buffer region beyond $\hat{x}/\delta > 0.25$ for the present database.  Therefore, our findings confirm that a substantial part of the internal layer remains in a non-equilibrium state with the local wall condition \citep[see also][]{Rao1974, antonia1972response, Shir1972}.

In the present set of experiments, a Preston tube is not tested as a method to measure wall-shear stress. However, since the typical diameter of a Preston tube is $O(1)$ mm which for this flow is approximately 30 wall units, we would expect similar errors from this device to those observed for the buffer layer fit shown in figure \ref{fig:Umean_HW_PTV}(\textit{a,c}). In short, calibration of Preston tubes are conducted under equilibrium smooth-walled conditions \citep[see][]{Patel1965}, and hence we would expect measurements with such a device to be compromised in the non-equilibrium buffer region flows occurring immediately downstream of a change in surface roughness.

\begin{figure*}
\vspace{5mm}
\centering
\includegraphics[trim = 0mm 0mm 0 0,scale = 0.7,clip]{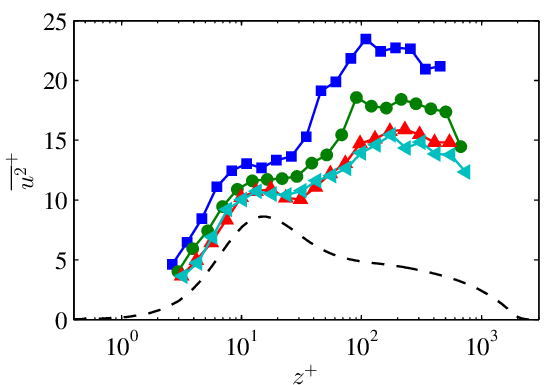}  \put(-175,130){(\textit{a})}  \put(-160,130){PTV  - buffer fit} 
\includegraphics[trim = 0mm 0mm 0 0,scale = 0.7,clip]{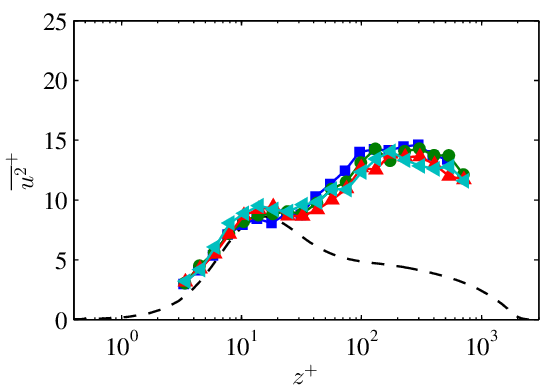}  \put(-175,130){(\textit{b})}  \put(-160,130){PTV - viscous sublayer fit} 
\caption{ Streamwise turbulence intensity, $\overline{u^2}^+$  from  the PTV experimental data at $Re_\tau \approx 4100$. Normalisation is by friction velocity, $U_\tau$ estimated from a fit to the  (\textit{a}) buffer region and  (\textit{b}) viscous sublayer. The black dashed line corresponds to a reference smooth walled boundary layer DNS database at $Re_\tau  \approx 2500$ \citep{Sillero2013} and the {\color{blue}$-\blacksquare-$}, {\color{green_m}$-\!\medbullet\!-$}, {\color{red}$-\!\blacktriangle\!-$} and {\color{cyan}$-\!\!\blacktriangleleft\!\!-$} symbols correspond to results at $\hat{x}/\delta=$ 0.11, 0.34, 0.68 and 1.0, respectively.}
\label{fig:Uvar_PTV}
\end{figure*}

Figure \ref{fig:Uvar_PTV} presents the streamwise turbulence intensities, $\overline{u^2}^+$, from the PTV database, where   \ref{fig:Uvar_PTV}(\textit{a}) and \ref{fig:Uvar_PTV}(\textit{b})   are normalised by  $U_\tau$ estimated from the buffer and viscous sublayer regions, respectively.  The results show that the $\overline{u^2}^+$ profile appears to be significantly altered by the inaccurate estimate of $\tau_{w}$ (hence $U_\tau$) based on a buffer fit immediately after the rough-to-smooth transition. For example, both profiles exhibit an energetic site which develops at a wall-normal location in close proximity to the `inner-peak' reported in equilibrium smooth-walled boundary layers \citep{Smits2011a}. However, \ref{fig:Uvar_PTV}(\textit{a})  reveals a sharp reduction in the magnitude of this energetic site with increasing $\hat{x}$, while \ref{fig:Uvar_PTV}(\textit{b})  exhibits only a subtle reduction in magnitude.  As a consequence, these shortcomings could compromise any attempts at establishing the appropriate scaling or modelling of the flow behaviour after a sudden change in surface conditions.  We note similar behaviour for the inner-scaled wall-normal turbulence intensity, $\overline{w^2}^+$, which is not reproduced here for brevity. 

\subsection{Skin-friction coefficient}

\begin{figure}
 \hspace*{5 mm}{
\centering
\includegraphics[trim = 0 0mm 0 0,scale = 0.7,clip]{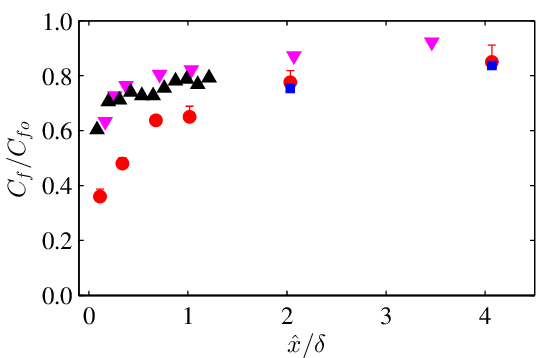}  }
\put(10,30){ \begin{tikzpicture}
\node[draw=white]{%
    \begin{tabular}{@{}r@{ }l@{}}
  \raisebox{0pt}{{\color{blue}$\blacksquare$}}&  Hotwire (log fit:  $3\sqrt{\delta^+} < z^+ < \delta^+_s$)~~~~~~~\\
\raisebox{0pt}{{\color{red}$\medbullet$}}&  Hotwire (buffer fit: $10 < z^+ < 30$)~~~~~~~\\
   \raisebox{0pt}{{\color{black}$\blacktriangle$}}&  PTV (viscous sublayer fit: $0 < z^+ < 4$)~~~~~~~ \\
      \raisebox{0pt}{{\color{magenta_m}$\blacktriangledown$}}&  OFI (direct)~~~~~~~ \\
    \end{tabular}};
\end{tikzpicture} }
\caption{ Skin-friction coefficient, $C_f$, estimates from the hotwire, OFI and PTV experimental data at $Re_\tau \approx 4100$. Normalisation is by $C_{fo}$, which correspond to the last measured magnitude of $C_f$ from the OFI database at $\hat{x}/\delta = 13.4$.}
\label{fig:Cf_HW_PTV}
\end{figure}
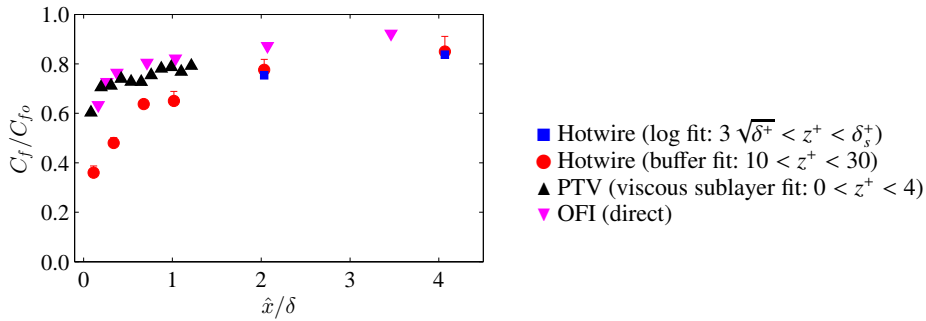

Figure \ref{fig:Cf_HW_PTV} compiles the skin-friction coefficient, $C_f$, downstream of the rough-to-smooth surface transition for all the current experimental databases. For PTV and OFI, $C_f$ is measured directly from the near-wall velocity gradient deep in the viscous sublayer, while the hotwire databases use either a buffer fit in the range $10 < z^+ < 30$ or a Clauser fit \citep{Clauser1954} in the expected log region. The Musker profile \citep{musker1979explicit} and composite velocity profile \citep{Chauhan2009} instead of the DNS data are also employed as the reference profile in the buffer region fit, and the scatter in the resulting $C_f$ is usually within $5\%$, as shown by the error bars in figure \ref{fig:Cf_HW_PTV}. Note that this error associated with the choice of the reference profile also presents in fully equilibrium smooth-wall boundary layers, therefore the fitted results should be interpreted with caution in general. For the Clauser fit we use constants $\kappa = 0.384$ and $B= 4.17$ in the range $3\sqrt{\delta^+} < z^+ < \delta_s^+$ (blue symbols). Here, the assumed upper limit of the logarithmic region $\delta_s^+$ is defined as $\min(0.15\delta^+,0.6\delta^+_i)$, where $\delta^+$ is the local viscous-scaled boundary layer thickness, and $\delta_i$ is the IBL thickness, defined as the `knee-point' in the $\overline{u^2}$ profile following \cite{efros2011development} and \cite{saito2014}. We prefer this method to identify $\delta_i$ as the distinction associated with the roughness change is more pronounced in $\overline{u^2}$ compared to $\overline{U}$ and less subject to small uncertainties in the measurement, resulting in a more robust estimation of $\delta_i$. The fit range is chosen in accordance with \cite{Marusic2013}, with an extra constraint to the upper limit as $0.6\delta^+_i$, as a different $U_{\tau}$ is expected above the IBL \citep{Elliott1958}. The coefficient 0.6 is empirically chosen to eliminate any `kink' in the mean velocity profile related to the internal boundary layer. Clauser fit results are not shown for $\hat{x}/\delta<2$ as $\delta_i^+$ is small in the immediate downstream of the surface transition, thus there is an insufficient number of data points to perform the fit.  Note that by performing a Clauser fit we do not imply the existence of a fully recovered log region in the immediate downstream of a rough-to-smooth transition. Our intention here is to demonstrate that, similar to the buffer fit as discussed in \S\ref{sec:exp_results}, if one takes the mean velocity data downstream of a rough-to-smooth transition and uses this to compute $U_\tau$ via a Clauser fit, an erroneous $U_{\tau}$ will result. 
It is worth emphasising that the spanwise variation in wall-shear stress can be appreciable immediately downstream of the rough-to-smooth change due to the effect of individual roughness elements \citep{MogengAFMC2018, wu2013turbulent}. Consequently, since the hotwire, PTV and OFI measurements are made at slightly different spanwise locations, any comparisons between them in the range $\hat{x}/\delta <0.4$ (corresponding to $\hat{x}/k_p <15$, approximately four times of the reattachment length as reported by \citet{wu2013turbulent} ) should be treated with caution. Nevertheless, if we define a recovery length $L$ as the downstream fetch where the local $C_f$ reaches, for example, $80\%$ of the full-recovery value $C_{f0}$, then $L = 0.8\delta$ for the OFI and PTV $C_f$ values, whereas $L = 2\delta$ for the buffer fit results. These results confirm that even beyond  $\hat{x}/\delta >0.4$ downstream of the transition the magnitude of $C_f$ is lower and exhibits a more gradual recovery as a function of $\hat{x}$ when estimated away from the wall (buffer and Clauser fits), as compared to estimates from closer to the wall (viscous sublayer) or at the wall (OFI). These discrepancies are likely to play a significant role in the wide range of recovery trends reported for $C_f$ in past works (see figure \ref{fig:IBL_sketch}(\textit{b}) ). In general, and to within experimental error, the OFI and PTV determined wall-shear stress are in close agreement. These observations will be revisited in \S \ref{sec:DNS_results}, where comparisons will be drawn to a  direct numerical simulation of a rough-to-smooth surface change in a wall-bounded flow. 

\section{Numerical experiment of a rough-to-smooth transition} \label{sec:DNS}

The results presented in the preceding discussions have highlighted that the accuracy of most `indirect' experimental techniques for estimating $\tau_w$ will be compromised in non-equilibrium conditions which persist in the near-wall and buffer region of the internal layer for surprisingly large distances downstream of the surface transition. To complement the experiments, a DNS database was generated and analysed to test for this behaviour.

\begin{figure}

\centering

\includegraphics[scale = 0.5,clip]{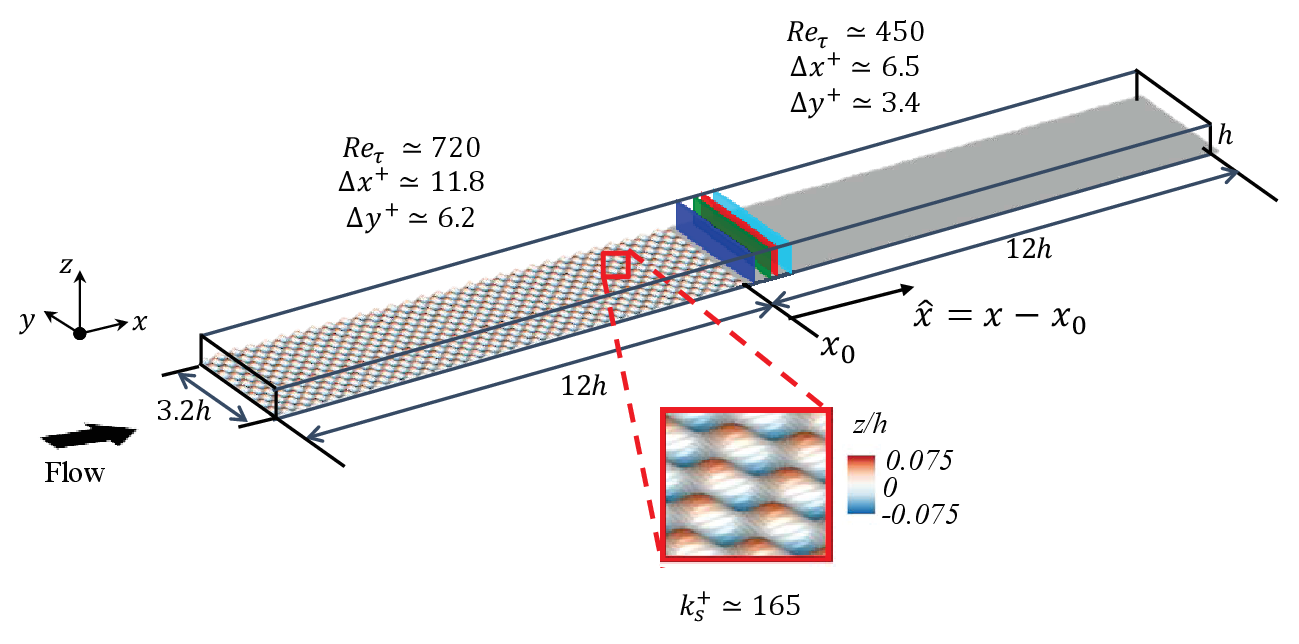}  
\put(-360,160){(\textit{a})} \\
\vspace{5mm}
\includegraphics[trim = 0 0mm 0 0,scale = 0.7]{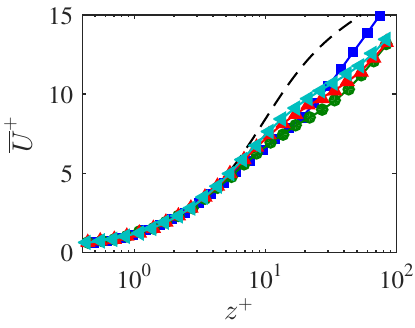}  
\put(-160,105){(\textit{b})} 

\caption{ (\textit{a}) The computational domain for the DNS database. The bottom surface is coloured according to the surface elevation relative to the location of the smooth wall plane. The inset shows a magnified view of the ``egg carton" roughness employed. The reported $Re_\tau$ on each patch corresponds to the recovered region and the wall parallel resolutions $\Updelta x^+$ and $\Updelta y^+$ are normalised by the asymptotic values of $U_\tau$ on each patch. (\textit{b}) Streamwise mean velocity, $\overline{U}$, normalised by the local $U_{\tau}$, from the DNS database at $Re_\tau \approx 590$. The black dashed line corresponds to a reference DNS channel flow database at $Re_\tau  = 934$ \citep{Alamo2004}. The {\color{blue}$-\blacksquare-$}, {\color{green_m}$-\!\medbullet\!-$},  {\color{red}$-\!\blacktriangle\!-$} and {\color{cyan}$-\!\blacktriangleleft\!-$} symbols correspond to $\hat{x}/h =$ 0.11, 0.34, 0.68 and 1.0 respectively, as shown by the wall-normal planes in (\textit{a}), where  $\hat{x}$ corresponds to the streamwise distance from the  rough-to-smooth transition.}
\label{fig:Umean_DNS}
\end{figure}

The DNS was performed using a well-validated fourth-order, finite-difference code~\citep{chung2014,chung2015} with an Immersed Boundary Method (IBM) used to implement the roughness~\citep{scotti2006, rouhi2018}. The open-channel computational domain for the present simulations spans $24h \times 3.2h \times h$ in the streamwise, spanwise and wall-normal directions, as shown in figure \ref{fig:Umean_DNS}(\textit{a}) . Periodic boundary conditions are applied in the streamwise and spanwise directions and a free-slip condition is employed at the top boundary. For $0 < x < 12h$ the bottom wall of the channel is a no-slip rough boundary, which then has an abrupt transition to a smooth wall no-slip boundary for $12h < x < 24h$. The rough patches are composed of an `egg carton' roughness~\citep{chan2015} with a roughness height of $0.056h$ and a roughness wavelength of $0.40h$, where $h$ corresponds to the channel height. Further, the midheight between the roughness crests and troughs is aligned with the smooth wall, corresponding to $\Updelta H/k_p = -0.5$. The flow at the end of the rough patch is in the fully rough regime with an equivalent sand-grain roughness of $k^+_s \approx 165$. The driving pressure-gradient is set such that the global Reynolds number is maintained at $Re_{\tau} = h U_{\tau_o} / \nu = 590$, where $U_{\tau_o}$ is the global friction velocity. The flow is fully resolved down to the roughness elements (approximately 24 points per roughness wavelength in the streamwise direction and 48 points in the spanwise direction) with no modelling assumptions. The wall-shear stress  $\tau_{w}$ over the smooth surface (and hence $U_\tau$) is computed from the gradient of the streamwise mean flow at the grid point closest to the wall (see Eq. \ref{eqn:utau_sublayer}), which is located below $z^+ < 0.5$ for the present case. Further details on the DNS database can be found in \citet{rouhi2018}.

\subsection{Results from a rough-to-smooth DNS database} \label{sec:DNS_results}

\begin{figure}
\centering
\vspace{5mm}
\includegraphics[trim = 0 0mm 0 0,scale = 0.7,clip]{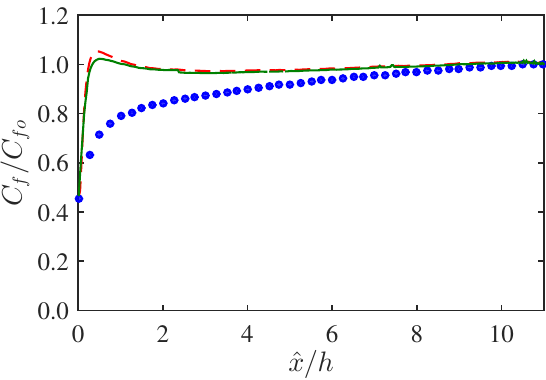}  
\put(-150,30){ \begin{tikzpicture}
\node[draw=white]{%
    \begin{tabular}{@{}r@{ }l@{}}
  \raisebox{0pt}{{\color{red}\textbf{$--$}}}& Direct~~~~~~~\\
   \raisebox{0pt}{{\color{green_m}\textbf{---}}}&  Viscous sublayer fit - $0 < z^+ < 4$ \\
      \raisebox{0pt}{{\color{blue}\textbf{$\bullet$}}}& Buffer fit - $10 < z^+ < 30$ \\
    \end{tabular}};
\end{tikzpicture} }
\put(-180,130){(\textit{a})}
\includegraphics[trim = 0 0mm 0 0,scale = 0.7]{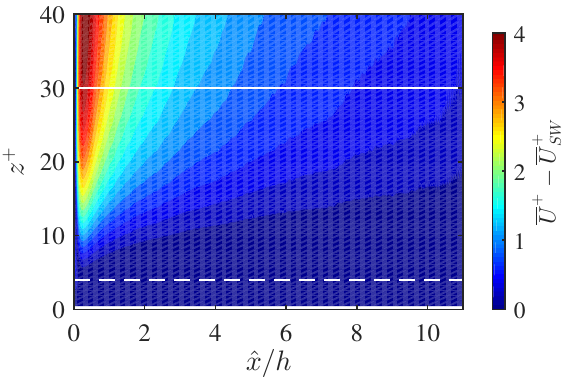}  
\put(-180,130){(\textit{b})}
\caption{ (\textit{a}) Skin-friction coefficient, $C_f$, estimates from the DNS database at $Re_\tau \approx 590$.  Normalisation is by $C_{fo}$, which corresponds to the last measured magnitude of $C_f$ for each case. (\textit{b}) Colour contours of the difference in the mean streamwise velocity ($\overline{U}_d^+ = \overline{U}^+-\overline{U}_{SW}^+$) immediately downstream of a rough-to-smooth transition relative to a reference smooth-walled open-channel flow, $\overline{U}_{SW}^+$, at a comparable $Re_\tau$. Results are  computed from DNS data where $U_\tau$ can be directly estimated from the velocity gradient at the wall. The white dashed and solid lines correspond to the upper limit of the viscous sublayer and buffer region, respectively.}
\label{fig:Cf_DNS}

\end{figure}
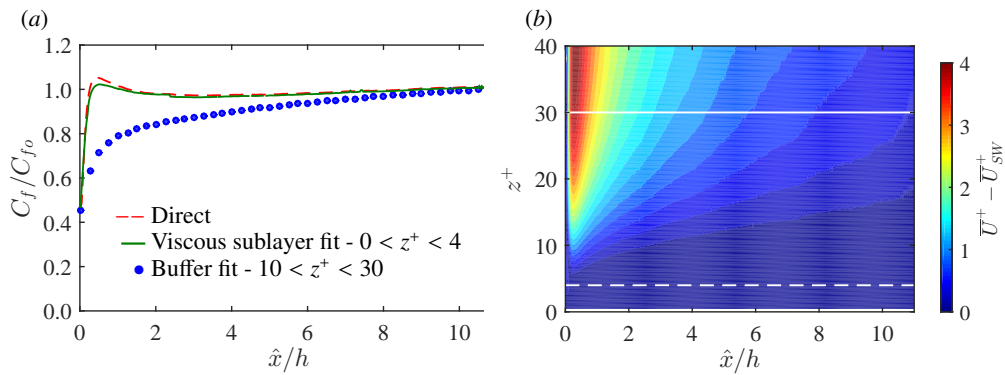

Figure \ref{fig:Umean_DNS}(\textit{b}) presents the inner-normalised streamwise mean velocity $\overline{U}^+$ from the DNS database at various locations downstream of the rough-to-smooth transition.  In the viscous sublayer ($z^+ \lesssim 4$) the results exhibit good agreement with the reference smooth-walled profile (dashed line, taken from \citet{Alamo2004} at $Re_\tau  = 934$). However, in the same manner as observed previously for the correctly scaled PTV experiments, the buffer region and beyond exhibits poor agreement. Note that the mean flow recovers to the equilibrium state monotonically in the experiments as shown in figure \ref{fig:Umean_HW_PTV}(\textit{b}), whereas in the simulation the mean velocity profile at $\hat{x}/h = 0.11$ (the blue curve) overshoots the rest three profiles further downstream. This discrepancy of the flow behaviour in the vicinity of the roughness transition can be attributed to the difference in the roughness height ($\delta/k_p \approx 45$ in the experiment versus $h/k_p \approx 9$ in the simulation). These results confirm that the buffer region requires a surprisingly long recovery length downstream of a rough-to-smooth transition to reach an equilibrium state that reflects the new smooth wall condition. As a consequence, any estimate of $\tau_{w}$ (hence $U_\tau$) obtained from the buffer region or above will be compromised. The extent of this discrepancy is highlighted by plotting the skin-friction coefficient $C_f$ calculated from various methods, downstream of the rough-to-smooth transition for the DNS data. These results are presented in figure \ref{fig:Cf_DNS}. The blue dotted curve shows the case where $\tau_{w}$ is estimated from the buffer region (fit in the range $10 \lesssim z^+ \lesssim 30$). This case exhibits a much longer and more gradual $C_f$ recovery as compared to the direct measure from the DNS (red dashed curve, obtained from the velocity gradient at the first off-wall grid cell).  On the other hand, when $C_f$ is estimated from the viscous sublayer region ($z^+\lesssim4$, the green curve of figure \ref{fig:Cf_DNS}(\textit{a}) ), we observe closer agreement to the direct measure from the DNS database. This is promising for experiments, where the viscous sublayer is certainly more accessible to measurements than the gradient at the wall (e.g. the PTV measurements presented previously). For the present case, the error between the viscous sublayer fit and the wall gradient measure drops from approximately from $5\%$ to $1\%$  as $\hat{x}/h$ increases from 0 to 2. These observations from the DNS data in figure \ref{fig:Cf_DNS} reconfirm the broad trends of  $C_f$ recovery for the various $\tau_w$ estimation techniques observed from the experiments (figure \ref{fig:Cf_HW_PTV}) and past works (figure \ref{fig:IBL_sketch}(\textit{b}) ), thus providing an explanation for some of the scatter observed. It is noted from a comparison of figure \ref{fig:Cf_DNS} with figure \ref{fig:Cf_HW_PTV}, that the buffer layer computed $C_f$ recovery following the rough-to-smooth transition in the DNS is quite different to that from the experiments, with the DNS indicating a slower recovery. This suggests that the DNS retains non-equilibrium effects in the buffer layer to a greater distance downstream of the rough-to-smooth transition than the experiments. It is noted that the DNS is at a much lower Reynolds number, with a much greater $k_p/\delta$ and is an open channel geometry, all of which would likely affect the recovery. Regardless, in the context of this study the overarching message is clear from both experiments and DNS: estimates of $C_f$ made further from the wall (in the buffer or log region) can be surprisingly inaccurate, even at several boundary layer thicknesses downstream of the transition.

We additionally note that the DNS data exhibit an overshoot of $C_f$ immediately downstream of the rough-to-smooth transition which is notably absent in the experiments (figure \ref{fig:Cf_HW_PTV}). This might be related to the difference of the step height $\Updelta H$ at the roughness transition, as the a greater down step is present in the experiment ($\Updelta H/ k_p = -1.5$) compared to the simulation ($\Updelta H/ k_p = -0.5$). Another possible factor for this behaviour might be associated with the lower $Re_{\tau}$ of the DNS database or the difference in geometry (appendix \ref{appA}). Further, the results from the DNS database also appear to exhibit a much slower recovery of the buffer region to the new wall conditions as a function of $h$ (or $\delta$) when compared to the experiments. This observation may be associated with the significantly larger $k_p/\delta$ ratio in the DNS databases compared to the experiments ($k_s/h \approx 0.2$ for the DNS, compared to $k_s/\delta \approx 0.04$ for the experiments). In any case, the broader trends from the DNS database confirm that any estimation of  $\tau_{w}$ made using the data above the viscous sublayer region is compromised for several $\delta$ downstream of a rough-to-smooth transition.

In order to quantify the rate of recovery of a wall-bounded flow to equilibrium conditions downstream of a rough-to-smooth surface change,  figure \ref{fig:Cf_DNS}(\textit{b})  presents colour contours of the difference in the streamwise mean flow, $\overline{U}_d^+$, after the rough-to-smooth transition relative to a fully developed smooth-walled flow, $\overline{U}_{SW}^+$, from a DNS database in the present study at matched $Re_\tau$. In this case, a direct measure of the velocity gradient at the wall and hence $\tau_{w}$ is available from both databases. For simplicity, comparisons are drawn for the same flow geometry (open-channel flow) to avoid the spatial growth of a turbulent boundary layer. The colour contours of $\overline{U}_d^+$ indicate an almost immediate recovery in the viscous sublayer region ($z^+\lesssim4$ indicated by the horizontal dashed line) to an equilibrium-state of a smooth-walled channel flow, while further from the wall, in the buffer region and beyond, larger discrepancies are present throughout the range  $0\!<\!\hat{x}/h\!<\!5$. These results confirm that (for a channel flow, and consistent with boundary layers) only beyond  $\hat{x}/h\gtrsim 5$, can we reliably employ diagnostic tools that operate in the buffer region to estimate $\tau_{w}$.

\section{Premultiplied energy spectrum} \label{spectra}

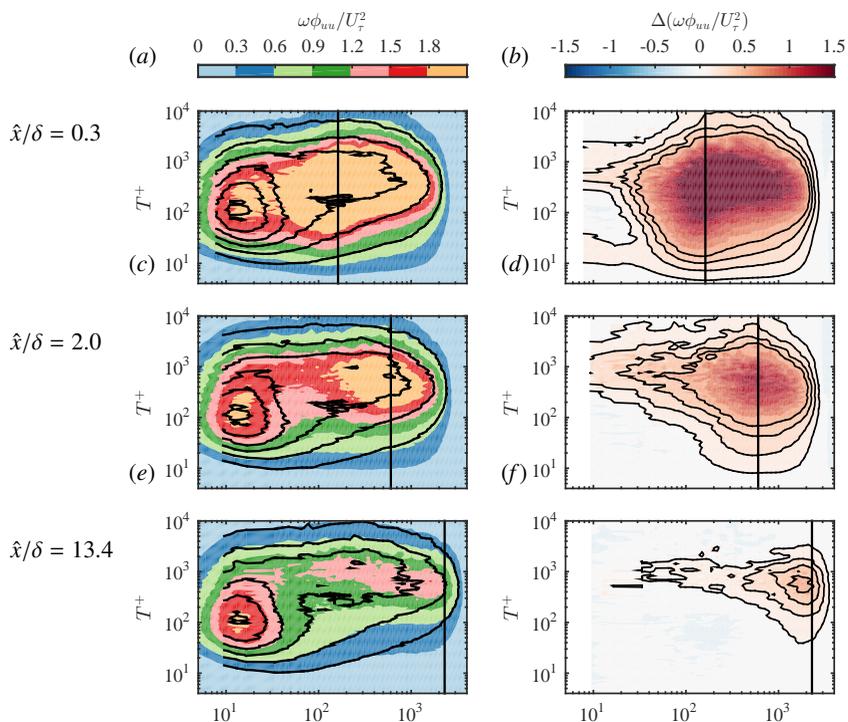
\begin{figure}
\centering
\vspace{10mm}
\includegraphics[trim = 0 7mm 0 0,scale = 0.85]{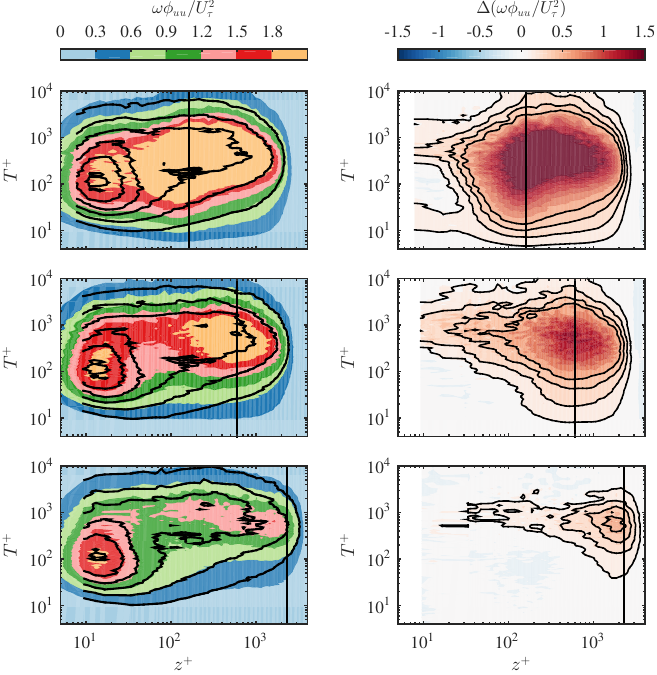}   \put(-270,240){(\textit{a})} \put(-270,162){(\textit{c})} \put(-270,84){(\textit{e})} \put(-130,240){(\textit{b})} \put(-130,162){(\textit{d})} \put(-130,84){(\textit{f})}
\put(-315,211){$\hat{x}/\delta = 0.3$}
\put(-315,133){$\hat{x}/\delta = 2.0$}
\put(-315,55){$\hat{x}/\delta = 13.4$}
\caption{Premultiplied energy spectra $\omega \phi_{uu}/U^2_{\tau}$ at (\textit{a}) $\hat{x}/\delta = 0.3$, (\textit{c}) 2.0 and (\textit{e}) 13.4. The coloured contour is the rough-to-smooth case, and the black contour lines are from a reference smooth-wall experiment at matched $Re_{\tau}$, with contour levels of $\omega \phi_{uu}/U^2_{\tau}$ = 0, 0.3, 0.6, 0.9, 1.2, 1.5, 1.8. (\textit{b, d}) and (\textit{f}) are the difference between the  rough-to-smooth case and the reference smooth case $ \Updelta (\omega\phi_{uu}/U_\tau^2)$ at streamwise locations corresponding to the left column. The four black contour lines indicate  0.15, 0.30, 0.45 and 0.60. The vertical black lines in all the figures represent the location of the IBL estimated from the turbulence intensity profile following  \cite{efros2011development} and \citet{saito2014}.}
\label{fig:spectra}
\end{figure}

From the preceding discussions, it is clear that the boundary layer gradually recovers to an equilibrium state of the new wall conditions after several boundary layer thicknesses downstream of the transition, yet it is not obvious which scales are responsible for this slow recovery. To provide a clearer picture of the recovery scale by scale, premultiplied energy spectra $\omega \phi_{uu}/U^2_{\tau}$ are shown in figure \ref{fig:spectra}, where $\omega = 2\pi /T$ is the angular frequency, $T$ is the time period (corresponding to the wavelength in spatial domain), $\phi_{uu}$ is the energy spectrum of the streamwise velocity fluctuation ($\int_0^{\infty} \phi_{uu} \mathrm{d} \omega = \overline{u^2}$), and $U_{\tau}$ is the friction velocity measured from the OFI experiments (see \S \ref{sec:exp_results}). The spectrograms presented are computed from hotwire time series data. Further, since the flow is heterogeneous in $x$, we refrain from converting the spectrum from temporal to the spatial domain, which has been shown to have limited accuracy in rough-walled flows \citep{squire2017applicability}. The colour contour maps in the left column are computed from the rough-to-smooth cases, while the solid contour lines represent a smooth-wall reference, which is obtained by interpolating the spectrum reported by \cite{marusic2015evolution} to the same $Re_{\tau}$ as the corresponding rough-to-smooth case. The length of the hotwire filament in viscous units is $l^+ \approx 17$ for the present rough-to-smooth case at $\hat{x}/\delta = 13.4$, while $l^+ \approx 24$ in the smooth-wall reference, leading to a $5\%$ more attenuation at the inner-site of the spectrum in the rough-to-smooth case compared to the reference \citep{Chin2010}. The results reveal clear evidence that the rough-wall structures are present beyond the IBL and are over-energised relative to the local $U_{\tau}$. In addition, there are signs that even within the IBL, the large-scale motions are over energised, providing further evidence that the IBL has not returned to equilibrium conditions. These results also show better agreement at smaller scales ($T^+<90$), particularly in the near wall region at larger  $\hat{x}$. A similar observation has also been made by \cite{ismail2018simulations} following a transition from transverse square ribs to a smooth wall in a channel flow DNS.

To further elucidate this behaviour, the right column shows the difference between the rough-to-smooth spectrum and the reference smooth-walled spectrum, defined as,
\begin{equation}
\Updelta (\omega\phi_{uu}/U_\tau^2) \equiv (\omega\phi_{uu}/U_\tau^2)_{R\rightarrow S} - (\omega\phi_{uu}/U_\tau^2)_{S}.
\end{equation}
These difference plots confirm that the energy distribution of the smaller scales recovers first, while the larger scales remain over-energised, reflecting the upstream rough wall condition. Interestingly, these over-energised large scales are not just restricted to the region above the IBL, but retain a footprint deep into the buffer region. These results suggest that the near-wall region recovering over the smooth surface will be subjected to a heightened degree of superposition and modulation from the over-energised large-scale events which retain the rough-wall upstream history \citep{Mathis2009}.

\subsection{An alternative method to estimate $U_{\tau}$}

The energy spectrum has revealed that the smaller energetic scales in the near-wall region appear to rapidly recover to equilibrium with the new smooth-walled surface. Based on this observation, we propose an alternative method to estimate $U_{\tau}$ for the flow downstream of a rough-to-smooth transition when no direct measurement at the wall or within the viscous sublayer is available. The essence of this method is to minimise the difference of the energy spectrum of the small scales in the near-wall region between the rough-to-smooth case and a smooth-wall reference dataset by adjusting the velocity scale $U_{\tau}$ for the rough-to-smooth case. There is some precedent for this approach in the literature for smooth wall canonical wall-bounded turbulent flows. \citet{Hutchins2009} have shown that over a range of Reynolds numbers, the energy in small scales appears to collapse on to a universal distribution when scaled by the local $U_{\tau}$. \citet{ganapathisubramani2018law} has shown that this universality is also persistent under the influence of freestream turbulence, and \cite{Monty2009} have observed small-scale universality between pipe, channel and boundary layer geometries. All of these cases suggest small-scale universality in the near-wall region, even in situations where we expect there to be large differences in the footprinting of the large-scales onto the near-wall region. 
\begin{figure}
\vspace{3mm}
\centering
\includegraphics[trim = 0 0mm 0 0,scale = 1,clip]{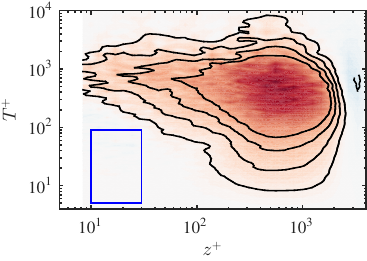}  
\put(-180,130){(\textit{a})}
\put(-100,30){$S$}
\put(-110,30){\line(5,1){10}}
\hspace{0.4cm}
\includegraphics[trim = 0 0mm 0 0,scale = 1,clip]{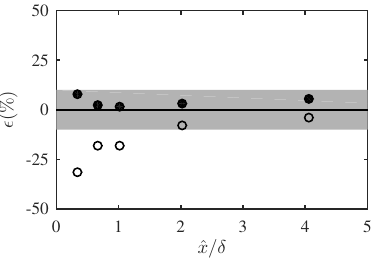}  
\put(-180,130){(\textit{b})}
\caption{(\textit{a}) Difference in premultiplied spectrum $\omega \phi_{uu}/U^2_{\tau}$ between the rough-to-smooth case at $\hat{x}/\delta = 2$ with estimated $U_{\tau}$ ($U_\tau$ is adjusted such that the integral of the difference across the blue rectangle region $S$ is minimum) and the smooth-wall reference. Contour levels are the same as in figure \ref{fig:spectra}(\textit{d}). (\textit{b}) $\epsilon = (C_f|_{M}-C_f|_{OFI})/C_f|_{OFI}$, where $M$ stands for buffer region fit (open symbols) or the spectrum fit (solid symbols). $\epsilon$ is the error relative to the OFI results. The shaded band covers $-10\%$ to $10\%$ on the vertical axis. Note that the data points at $\hat{x}/\delta = 0.11$ are not shown in the figure as they fall beyond the current axis limit.}
\label{fig:utau_surro}
\end{figure}

For the present case, a rectangular region $S$ (marked in blue on figure \ref{fig:utau_surro}(\textit{a}) ) of energetic scales in the near-wall region is chosen that is bounded by the limits $10<z^+<30$ and $5<T^+<90$. These bounds are chosen empirically based on the vanishing $\Updelta(\omega\phi_{uu}/U_\tau^2)$ observed in this region from figure \ref{fig:spectra}.  The difference between the viscous scaled energy spectra for the rough-to-smooth and the smooth case $\Updelta(\omega\phi_{uu}/U_\tau^2)$, is then minimised across this region by varying $U_\tau$ for the rough-to-smooth case. Figure \ref{fig:utau_surro}(\textit{a})  shows an example where $U_\tau|_{R\rightarrow S}$ has been optimised in this manner, to give the minimum $\Updelta(\omega\phi_{uu}/U_\tau^2)$ within the rectangular region $S$. To test the efficacy of this method of determining $U_\tau$,   figure \ref{fig:utau_surro}(\textit{b})  presents the relative error $\epsilon$ in $C_f$ obtained using the spectrum fit (solid symbols) and buffer region fit (open symbols) as compared to the OFI results. The results show that the spectrum fit, although still subject to error, provides a better estimate of $U_{\tau}$ immediately downstream of the rough-to-smooth transition compared to methods that purely rely on the mean streamwise velocity in the buffer region. We note, the precise nature of dependence of the bounds of $S$ on $Re_{\tau}$, $k^+_s$ and other flow parameters remain to be examined by performing more experiments covering a broader range of conditions in future works.

\section{Summary and conclusions} \label{sec:conclusions}

This work presents a systematic study on estimating the wall-shear stress, $\tau_w$, after a sudden change in surface conditions from a rough-to-smooth wall.  To this end, a unique collection of experimental and numerical databases are examined offering access to both `direct' and `indirect' measures of $\tau_w$. Our experimental results reveal that the mean flow within the buffer region (defined as $10 < z^+ < 30$) only recovers to an equilibrium state with the new local smooth-wall conditions after approximately 5 boundary layer thicknesses downstream of the rough-to-smooth transition. Based on these findings, `indirect' techniques that only have access to velocity information above the viscous sublayer are shown to consistently underestimate the magnitude of $\tau_w$ immediately downstream of a rough-to-smooth transition. This discrepancy, in turn, can give the erroneous impression of a longer recovery length of  $C_f$ to the new wall conditions and is likely to be responsible for the wide range of recovery trends reported for $C_f$ following a rough-to-smooth transition. More specifically, the further from the wall that $C_f$ is inferred from the velocity profile, the greater the underestimation of $C_f$, and the greater the recovery length.
 
To complement the experimental databases and further confirm our findings, a DNS database with comparable flow conditions and access to a direct measure of $\tau_w$ is employed. These data lead to similar conclusions, indicating that in the range $0\!<\!\hat{x}/h\!<\!5$, an accurate estimate of the wall-shear stress estimates of $\tau_w$ can only be obtained in the viscous region ($z^+\lesssim 4$). More specifically, diagnostic tools that operate in the buffer region are likely to provide a reliable estimate of the wall-shear stress only beyond $ \hat{x}/h \gtrsim 5$ downstream of a rough-to-smooth transition. 

Through an analysis of the energy spectra we observe that the smaller energetic scales ($T^+<90$) in the buffer region adjust to the new wall condition over a relatively short recovery ($\hat{x}/\delta\!\lesssim\!1$). Conversely, the large-scale motions ($T^+>90$), which are over-energised (relative to the new smooth wall boundary condition) retain a strong footprint in the IBL, extending deep into the buffer region. Based on the observation that the small scales attain a universal form over relatively short recovery distances, an alternative approach to estimate the wall-shear stress from the premultiplied energy spectra is proposed when no direct measurement of the wall-shear stress is available. The results reveal improved performance relative to more conventional techniques that are based purely on the mean velocity profile in the buffer region.

\section*{Acknowledgements}
This research was partially supported under the Australian Research Council’s Discovery Projects funding scheme (project DP160103619). This research was also supported by resources provided by the Pawsey Supercomputing Centre with funding from the Australian Government and the Government of Western Australia and by the National Computing Infrastructure (NCI), which is supported by the Australian Government. 

\appendix
\section{$C_f$ data in literature with a direct measure of $\tau_{w}$}\label{appA}
\begin{figure}
\centering
\vspace{5mm}
\includegraphics[scale = 0.8]{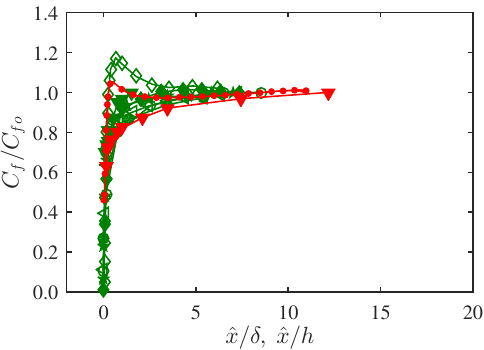} 
\caption{Revisit of the $C_f$ data from literature as shown in figure \ref{fig:IBL_sketch}(\textit{b}) . Only the datasets with a direct measurement of the wall-shear stress are shown, and symbols are the same as in table \ref{tab:DataSumm}. OFI and DNS results from the current study are represented by {\color{red}$\blacktriangledown$} and {\color{red}$\bullet$}, respectively.
}
\label{fig:Appen_Cf}
\end{figure}

In this paper, we have highlighted that the scatter for the recovery of $C_f$ after a rough-to-smooth transition appears to be partly due to the measurement techniques employed. However, the recovery of $C_f$ can be affected by a number of factors (see \S \ref{sec:intro}) including the Reynolds number, flow geometry (boundary layer or channel and pipe) and the roughness geometry. In order to examine some of these additional factors, figure \ref{fig:Appen_Cf} presents a subset of the datasets previously shown in figure \ref{fig:IBL_sketch}(\textit{b})  and table \ref{tab:DataSumm} that have access to a direct measure of the wall-shear stress. Consequently, we are limited to comparing data from the present study and the data of  \cite{Chamorro2009} and \cite{ismail2018effect, ismail2018simulations}. Although limited by available data, figure \ref{fig:Appen_Cf} suggests the overshoot in the current DNS database and the lowest $Re$ database from \cite{ismail2018simulations} ({\color{green_m}$ \lozenge$}) might be a low Reynolds number effect. Certainly, the higher $Re$ data from \cite{ismail2018simulations} ({\color{green_m} $\blacklozenge$}, {\color{green_m} $\lhd$} and {\color{green_m} $\star$}) do not exhibit this overshoot. Further data are required to confirm this tentative observation. In addition, the boundary layer data shown in figure \ref{fig:Appen_Cf}  (OFI in the present study and near-wall hotwire in \cite{Chamorro2009}) reveal a substantial difference in the recovery length.  However, from table \ref{tab:DataSumm} it is noted that the rough-to-smooth case of \cite{Chamorro2009} had a much higher $Re_{\tau}$ and $k_s^+$ than the current experimental study, which may suggest further influencing factors. Furthermore, datasets with a direct measure of $\tau_w$ are dominated by DNS studies, and they are mostly conducted with a channel configuration at low Reynolds numbers with high $k_s/h$ values. These tendencies may also bias the comparison. To answer these questions future works over a wide range of $Re$, roughness parameters and flow geometries with a direct measure of wall-shear stress are necessary.

\bibliographystyle{jfm}

\bibliography{MAIN_BIB}

\end{document}